\documentclass[runningheads]{svmult}
\usepackage{makeidx}   
\usepackage{graphicx}  
\usepackage{subeqnar}  
\usepackage{multicol}  
\usepackage{cropmark} 
\usepackage{physprbb}  
\begin{document}
\title*{Self-tuning solutions of the cosmological constant}

\author{Jihn E. Kim
}

\institute{Division of Physics, Seoul National University, Seoul
151-747, Korea}

\maketitle

\begin{abstract}
The self-tuning solutions of the cosmological constant is
reviewed, with the emphasis on the recent attempts in extra
dimensional gravity with a brane boundary.
\end{abstract}

Since 1975, the hierarchy problems become the most challenging
problems in particle physics:\\

\indent (i) the gauge hierarchy problem\cite{gauge},

\indent (ii) the cosmological constant problem\cite{ccp},

\indent (iii) the strong CP problem\cite{strongcp}, and

\indent (iv) the $\mu$ problem\cite{mu}, etc.\\

The most severe one among these is the cosmological constant
problem in that the hierarchy is the largest and there seems to be
no accepted solution in this problem, while the other hierarchy
problems leads to some plausible solutions such as supersymmetry,
axion, and the introduction of the hidden sector. This
cosmological constant problem attracted an intense scrutiny from
many famous physicists
\cite{ccp,weinberg,hawking,witten,weinberg1,coleman}.

Gravity is believed to be described by the metric $g_{\mu\nu}$
which defines the distance between two points. If the space is
flat, the pseudo-Pythagoras rule gives
\begin{eqnarray}\label{flat}
ds^2=-dt^2+(d{\bf x})^2=\eta_{\mu\nu}dx^\mu dx^\nu\ \ {\rm flat\
space. }\nonumber
\end{eqnarray}
If we change the flat metric $\eta_{\mu\nu}$ to a curved one
$g_{\mu\nu}$, we have the gravity equation,
\begin{equation}
R_{\mu\nu}-\frac12 Rg_{\mu\nu}=8\pi G T_{\mu\nu}
\end{equation}
where $8\pi G$ is the inverse Planck mass squared $1/M^2_P$ and
$T_{\mu\nu}$ is the energy momentum tensor constructed from the
fields except $g_{\mu\nu}$. The left-hand side vanishes when the
space-time is flat. If the right-hand side is non-vanishing, the
space-time cannot be flat. Under the cosmological principles, we
can solve this equation for the evolving universe. In 1910's, the
universe seemed to be not evolving, and in 1917 Einstein
introduced a compensating term to make the universe static. The
only possible 2nd rank tensor is $\Lambda g_{\mu\nu}$ which is
called the cosmological constant term. But, after the discovery of
the expanding universe by Hubble in 1929, the cosmological
constant lost the initial motivation for its introduction.

The flat space metric (\ref{flat}) satisfies the Pythagoras rule.
But curved spaces do not satisfy the Pythagoras rule. With
non-vanishing cosmological constant $\Lambda$, the space-time is
curved. For example, the de Sitter space metric
\begin{equation}\label{deSitter}
ds^2=-dt^2+e^{2\sqrt{\Lambda} t}(d{\bf x})^2
\end{equation}
is certainly curved. Thus, in a universe with a cosmological
constant, one can in principle measure the curvature. Its upper
bound is known to be extremely small,
$10^{-55}$~cm$^{-2}$\cite{zel2}. Therefore, it has been believed
for a long time that the cosmological constant is better to be
zero. But the recent Type 1A supernovae data suggests the vacuum
energy of order (0.003 eV)$^4$\cite{type1a}, which adds another
difficulty in understanding the cosmological constant problem. But
it is reasonable to view this problem as depicted in Fig. 1.
Namely, the true vacuum has zero vacuum energy, but the present
universe has not reached its absolute minimum yet. This situation
is generally studied as quintessence\cite{quintessence} with a
slowly varying dark energy with $\omega\le -\frac13$. Of course,
there is a possibility that the potential energy at the true
minimum is (0.003 eV)$^4$. Even for this case, one may need to
understand how the universe with a vanishing cosmological constant
is obtained.

\begin{figure}[b]
\begin{center}
\includegraphics[width=.95\textwidth]{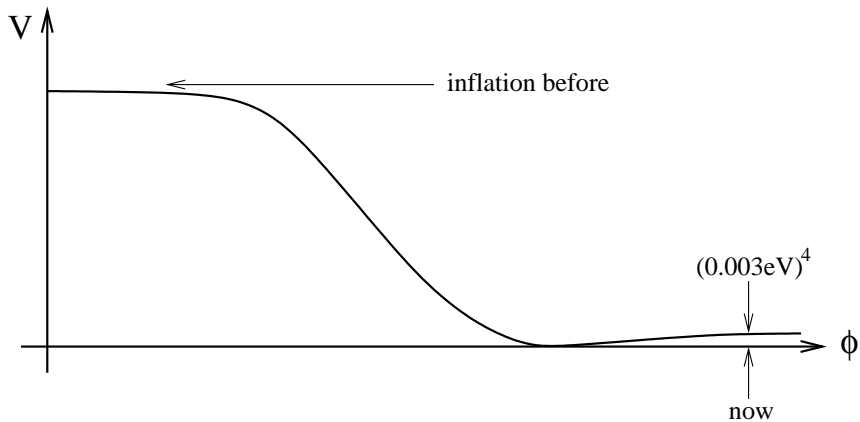}
\end{center}
\caption[]{A schematic view of the vacuum energy. The direction of
$\phi$ is through multi-dimensional fields, not into a single
scalar.} \label{eps1}
\end{figure}

One can see the cosmological constant problem very easily from the
action, guessed from the Einstein equation,
\begin{equation}
S=\int d^4x\sqrt{-g}\left(-\frac{M^2}{2}R\right)
\end{equation}
where $R$ is the Ricci scalar. But the general covariance allows
not only $R$, but also a constant and higher order terms of $R$,
etc. In particular, adding a constant is perfectly good,
\begin{equation}\label{ccaction}
S=\int d^4x\sqrt{-g}\left(-\frac{M^2}{2}R-V_0\right),
\end{equation}
from which we obtain the equation supplied with a cosmological
constant, $\Lambda=8\pi GV_0=V_0/M_P^2$,
\begin{equation}
R_{\mu\nu}-\frac12 Rg_{\mu\nu}-8\pi GV_0 g_{\mu\nu}=8\pi G
T_{\mu\nu}.
\end{equation}
The difficulty with the cosmological constant problem is that
there seems to be no symmetry forbidding the constant term in the
action. An obvious symmetry can be the scale invariance, but it is
badly broken. Even the electroweak scale introduces a mass scale
of $10^{56}$ orders larger than the observed upper bound on the
cosmological constant.

This cosmological constant problem surfaced as a very serious one
when the spontaneous symmetry breaking in electroweak theory was
extensively studied in particle physics\cite{ccp}. Here, the Higgs
potential looks as shown in Fig. 2.

\begin{figure}[b]
\begin{center}
\includegraphics[width=.95\textwidth]{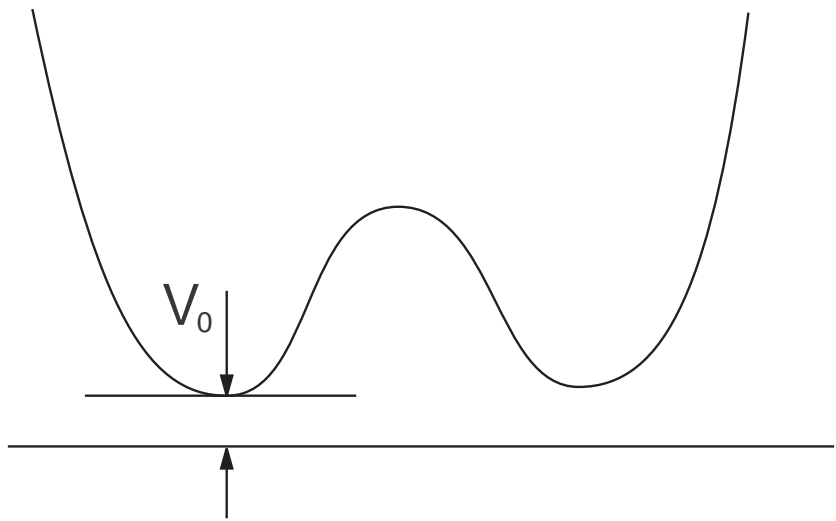}
\end{center}
\caption[]{The Higgs potential.} \label{eps2}
\end{figure}

Thus, here the vacuum energy itself is the cosmological constant.
The constant in Eq.~(\ref{ccaction}) is viewed as the contribution
from the Higgs potential. Since this Veltman's
observation\cite{ccp}, the cosmological constant became the most
difficult hierarchy problem in particle physics. As seen above,
from the action it is quite natural to allow a cosmological
constant. One may not prefer to choose a flat universe as a
boundary condition for the equation of motion, but prefers the
theory itself to choose a flat universe out of numerous
possibilities. It seems that the logic for introducing a
cosmological constant from the action is more fundamental than
Einstein's reasoning for a static universe.

The Einstein equation can be solved with the ansatze of a flat
universe, a de Sitter space universe, or an anti de Sitter space
universe. If a solution exists for a flat ansatz, then it gives a
vanishing cosmological constant. If a solution exists for a de
Sitter space ansatz, then it gives a positive cosmological
constant, etc. It is like solving the equation of motion to seek
the minimum of the potential. Note that in 4D for a nonzero
cosmological constant, a flat space solution is not possible. To
have a flat universe in a 4D theory, one must fine-tune $\Lambda$
at zero, which is the fine-tuning problem of the cosmological
constant.

But our 4D may come from higher dimensions. In a higher
dimensional theory, e.g. in 5D, we have a 5D cosmological
constant. But we live in 4D, and it is perfectly allowable if the
effective cosmological constant to a 4D observer is zero. In this
regards, the Randall-Sundrum models\cite{rs1,rs2} are very
interesting. These models start with a negative 5D cosmological
constant $\Lambda_b$, but look for 4D flat solutions,
\begin{equation}
ds^2=\beta(y)^2(-dt^2+d{\bf x}^2)+dy^2
\end{equation}
where $y$ is the fifth coordinate and $\beta$ is called $\lq$warp
factor'. The first question is, $\lq\lq$Does this ansatz allow a
solution?" For the Randall-Sundrum model II(RS-II)\cite{rs2}, we
consider a 4-brane located at $y=0$. At the brane matter can be
present, in particular a 4D cosmological constant or the brane
tension $\Lambda_1$. With a $Z_2$ ansatz a solution for $\beta(y)$
satisfies $\beta(y)=\beta(-y)$. Indeed, these ansatze allow a
solution for $\beta(y)$ which is expotentially suppressed at large
$|y|$. Even if this theory is a 5D theory, the effect of the deep
bulk($|y|\gg 1$) is negligible and can be considered as a good
effective 4D theory after integrating out with respect to $y.$ But
a flat solution is possible with the condition between parameters,
\begin{equation}\label{rsfine}
k_b=k_1,\ \ k_b=\sqrt{-\frac{\Lambda_b}{6M^3}},\ \
k_1=\frac{\Lambda_1}{6M}
\end{equation}
But this fine-tuning condition can be thought as an improvement
compared to the 4D case since we started with an arbitrary
negative bulk cosmological constant. It seems that the higher
dimensional theories may lead to a solution of the cosmological
constant problem.

Previously, there were several attempts toward a solution of the
cosmological constant problem, notably by Hawking, Witten,
Weinberg, and Coleman, under the name of probabilistic
interpretation in Euclidian quantum gravity\cite{hawking},
boundary of different phases\cite{witten}, anthropic
solution\cite{weinberg1}, and wormholes\cite{coleman}. For
example, the argument of anthropic principle is the following.
Life evolution is not very much affected by the existence of the
cosmological constant. But galaxy formation may be hindered if the
cosmological constant is too large. Weinberg obtained a bound for
the condensation of matter: $\rho<550\rho_c$ where $\rho_c$ is the
critical energy density. In this case, one needs a fine-tuning of
1 out of 1000, which is a big improvement. The anthropic solution
seems to work. But the anthropic solution needs a multi universe
scenario so that out of multiple universes, somewhere the
hospitable universe may exist.

In this talk, I review the self-tuning solution. I distinguish the
earlier version and the recent version. \vskip 0.5cm

\noindent {\bf Weak self-tuning solution}

If a solution for the flat space exists in the given theory, then
it is called the self-tuning solution or the undetermined
integration constant solution. For a specific value of the
undetermined integration constant $c$, a flat space is possible.
For example, Witten used the field strength $H_{\mu\nu\rho\sigma}$
of third rank antisymmetric tensor field. In this case, indeed
there is an undetermined integration constant $c$, and the vacuum
energy is proportional to $c^2$. This undetermined integration
constant can be chosen to make the space flat. Witten\cite{witten}
argued that probably the boundary of different phases is chosen.
The flat space is the boundary of the de Sitter and anti de Sitter
spaces. However, once $c$ is determined, there is no more free
parameter to adjust when a phase transition adds a further vacuum
energy. So in 4D this does not work, which has the root that
$H_{\mu\nu\rho\sigma}$ is not a dynamical field in 4D. Hawking
used the same field but his argument is a probabilistic one in
Euclidian quantum gravity. We will discuss more about Hawking's
idea later. \vskip 0.5cm

\begin{figure}[b]
\begin{center}
\includegraphics[width=.95\textwidth]{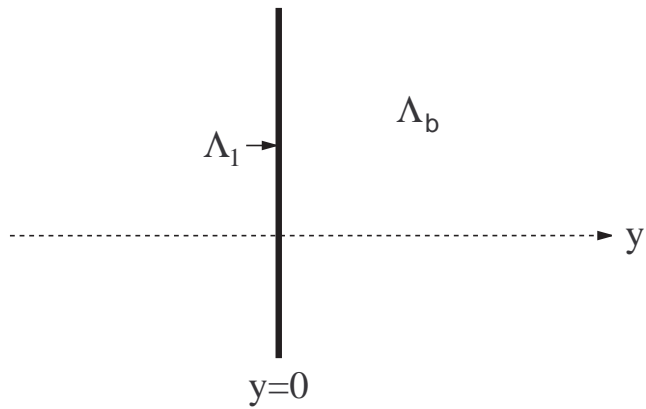}
\end{center}
\caption[]{The RS-II model.} \label{eps3}
\end{figure}

\noindent {\bf Strong self-tuning solution}

As seen above, the weak self-tuning solution does not care whether
there exist de Sitter or anti de Sitter space solutions. Choosing
the flat space solution came from another principle. Recently, a
more restricted class of self-tuning solutions have been
attempted\cite{kachru}. It requires that there is no de Sitter and
anti de Sitter space solutions except the flat one. Recently, this
strong self-tuning solution has attracted a great deal of
attention. In particular, the RS type models\cite{rs1,rs2,GB},
using a negative but nonzero bulk cosmological constant and
nonzero brane tension(s), allow 4D flat space solutions with
fine-tuning(s). Therefore, the RS models is a good playground for
the self-tuning solutions, starting with non-vanishing 5D vacuum
parameters.

Firstly, let us observe the key points of the RS type models in
terms of the RS-II model. The RS-II\cite{rs2} model is an
alternative to compactification. The fifth dimension $y$ is not
compactified, but one can obtain an effective 4D theory due to the
localization of gravity near $y=0$. It can be studied with the
following 5D Lagrangian,
\begin{equation}
{\cal L}=\frac{M^3}{2}(R-\Lambda_b)+({\cal L}_{\rm
matter}-\Lambda_1)\delta(y)
\end{equation}
where $M$ is the 5D fundamental scale, $\Lambda_b$ is the bulk
cosmological constant, and $\Lambda_1$ is the brane tension. It is
depicted in Fig. 3.

With the flat space ansatz, $ds^2=\beta(y)\eta_{\mu\nu} x^\mu
x^\nu+dy^2$, one can obtain a solution
\begin{equation}
\beta(y)=\beta_0 e^{-k|y|},\ \ \ k=\sqrt{-\frac{\Lambda_b}{6M^3}}.
\end{equation}
Thus, only the AdS bulk allows the solution. If there are more
branes, there are more conditions to satisfy toward a flat space
solution since there are more brane tensions. Thus, the RS-II
model is the easiest one(with only one brane tension) for a
self-tuning solution.

The strong self-tuning solution attracted a great deal of interest
because of a possibility to realize such a scheme\cite{kachru}.
They tried a Lagrangian with a dilaton-like field $\phi$,
\begin{equation}\label{kach}
{\cal L}=R-\Lambda^{a\phi}-\frac43(\nabla\phi)^2 -Ve^{b\phi}
\delta(y),
\end{equation}
where $a,V,$ and $b$ are constants. We set $M^3/2=1.$ It is a
RS-II type model with a massless scalar field in the bulk. This
scalar $\phi$ interacts with the brane tension. The relevant
equations of motion are
\begin{eqnarray}
{\rm dilation}\ &:& \frac83 \phi^{\prime\prime}+\frac{32}{3}
A^\prime\phi^\prime-a\Lambda e^{a\phi}-bV\delta(y)e^{b\phi}
=0\nonumber\\
{\rm (55)\ component}\ &:& 6(A^\prime)^2-\frac23(\phi^\prime)^2
+\frac12\Lambda e^{a\phi}=0\nonumber\\
{\rm (55),}\ (\mu\nu)\ &:& 3A^{\prime\prime}+ \frac43(\phi^\prime)
^2+\frac12 e^{b\phi}V\delta(y)=0\nonumber\\
\end{eqnarray}
where $A=\frac12\ln \beta(y)$ is the metric part with the 4D flat
space ansatz. Indeed, one can find a solution with this flat space
ansatz, for the relations $A^\prime=\alpha\phi^\prime$ and
$\Lambda=0$. The solution is
\begin{equation}
\phi=\pm \frac34\ln|\frac43 y+c|+d,\ \ \mbox{for the case of }
\alpha=\pm\frac13.
\end{equation}
There are two types for this function: one without singularity and
the other with a singularity at $y_c\equiv-\frac34 c$. The
non-singluar solution diverges logarithmically at large $|y|$ and
hence the localization of gravity near the brane is not realized.
For the singular solution, the singular point $y_c$ is the naked
singularity. But we cannot ignore the space $|y|>y_c$. Even if the
bulk for a given $y$ is a flat 4D, the effective theory must know
the whole $y$ space, including the boundary term if such a term is
present. In either case, the solution is not complete. For the
non-singular solution F$\ddot{\rm o}$rste {\it et
al.}\cite{nilles} tried to cure the singularity by inserting
another brane at $y=y_c$. Then a flat 4D solution is possible. But
there must be one fine-tuning. One more brane introduces one more
parameter(the brane tension), and for a specific value of this
additional brane tension the space is flat. Certainly, for another
value for this tension the solution is not flat. Therefore, it is
fair to say that we have not obtained a strong self-tuning
solution yet. \vskip 0.5cm

\noindent {\bf A weak self-tuning solution with $1/H^2$}
\\

For a self-tuning solution, it is a necessity to introduce
additional fields. In the RS-II model, one can guess that it is
better to introduce a massless scalar field since it must affect
the whole bulk. Indeed, Kachru {\it et al.}\cite{kachru} tried a
massless bulk scalar. It may be better if this massless scalar
arises from the symmetry argument. Indeed, in $(4+n)$ space-time
an antisymmetric tensor gauge field with $(2+n)$ indices gives a
massless scalar. Therefore, in 5D RS-II type models, let us
introduce $A_{MNP},\ (M,N,P=0,1,2,3,5)$. Its field strength is a
4-form field $H_{MNPQ}$. Under the gauge transformation
$$
A_{MNP}\rightarrow A_{MNP}+\partial_{[M}\lambda_{NP]},
$$
the 4-form field $H$ is invariant. There will remain a $U(1)$
gauge symmetry(but in fact global since there is only one
component of the dynamical field) with one massless pseudoscalar
$a$, $\partial_M a\propto \epsilon_{MNPQR}H^{NPQR}$. It remains to
be seen whether there exists a self-tuning solution.

Let us consider the simplest kinetic energy term,
\begin{equation}
S=\int d^5x\sqrt{-G}\left[\frac{M^3}{2}R-\frac{M}{2\cdot
4!}H^2-\Lambda_b \right] +\int d^4x\sqrt{-g}(-\Lambda_1)
\end{equation}
where $H^2=H_{MNPQ}H^{MNPQ}$, and $G$ and $g$ are the determinants
of the 5D and 4D metrics, respectively. The brane with the brane
tension $\Lambda_1$ is located at $y=0$. We have the following
ansatz,
\begin{eqnarray}
\mbox{Flat 4D}\ &:&\ ds^2=\beta(y)^2\eta_{\mu\nu}dx^\mu dx^\nu+dy^2
\nonumber\\
\mbox{4D chosen}\ &:&
H_{\mu\nu\rho\sigma}=\frac{\epsilon_{\mu\nu\rho\sigma}}{\sqrt{-G}
n(y)}\nonumber
\end{eqnarray}
where $\mu,\cdots,$ are the 4D indices. In the remainder of this
talk we work with the unit system $M=1$ unless the parameter is
explicitly inserted. The $H$ equation, the (55) and $(\mu\nu)$
component Einstein equations are solved to give the following
solutions:
\begin{eqnarray}
\Lambda_b<0 &:& \beta(|y|)=(a/k)^{1/4}[\pm \sinh(4k|y|+c)]^{1/4}
\nonumber\\
\Lambda_b>0 &:& \beta(|y|)=(a/k)^{1/4}[\sin(4k|y|+c')]^{1/4}\nonumber\\
\Lambda_b=0 &:&
\beta(|y|)=[|4a|y|+c^{\prime\prime}|]^{1/4}\nonumber
\end{eqnarray}
For a localizable(near $y=0$) metric, there exists a singularity
at $|y|=-c/4k,$ etc. Thus another brane is necessary and needs a
fine-tuning. We encounter a similar problem as in the case of
Kachru {\it et al.}\cite{kachru}. It is worth commenting the de
Sitter space solution, for $\Lambda_b>0$. It is a periodic
solution as shown in Fig. 4.

\begin{figure}[b]
\begin{center}
\includegraphics[width=.95\textwidth]{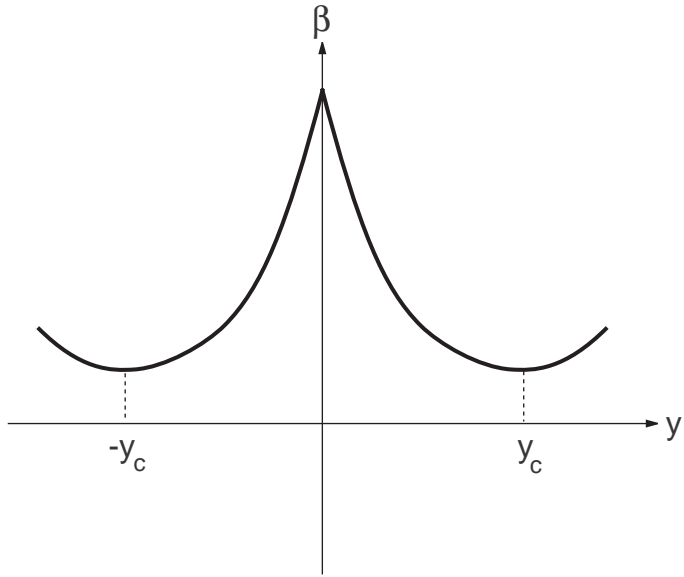}
\end{center}
\caption[]{The solution with $H^2$ in the bulk de Sitter space.}
\label{eps4}
\end{figure}

Since $\beta'=0$ at $y=\pm y_c$, we can restrict to the subspace
$|y|\le y_c$. Then the boundary conditions at $y=0$ and $y_c$
relate the parameters
$$
c=\cot^{-1} (k_1/k),\ \ \ c=4ky_c-\cot^{-1}(k^2/k),
$$
where $c$ is an integration constant determined by the brane
tension and the bulk cosmological constant. Then, if $y_c$ behaves
like an undetermined integration constant, we can achieve a
self-tuning solution. But $y_c$ is the vacuum expectation
value(VEV) of the radion field $g_{55}$ which cannot be a strictly
massless Goldstone boson. If it were massless, it will serve to
the long range gravitational interaction and hence give a
different result from the prediction of the Einstein gravity for
the light bending experiments. It will obtain mass, and hence
$y_c$ is not a free parameter but is fixed. Then the second
equation is nothing but a fine-tuning condition\cite{lee1}.

But we find a self-tuning solution with $1/H^2$\cite{kkl}. We
present this solution here, not because it has a meaningful
kinetic energy term but because to show that there exists a
solution. At low energy, we may treat it as an effective
interaction term. Setting $M=1$, the action is
\begin{equation}\label{H-2}
S=\int d^4x\int dy\left[\frac12 R+\frac{2\cdot 4!}{H^2}-\Lambda_b
-\Lambda_1\delta(y) \right]
\end{equation}
where $-\Lambda_1$ can be interpreted as the VEV of the matter
Lagrangian at the brane. At low energy, this Lagrangian makes
sense only if $H^2$ develops a VEV. Again, we impose the following
ansatze
\begin{eqnarray}\label{ansatz}
\mbox{Flat 4D}\ &:&\ ds^2=\beta(y)^2\eta_{\mu\nu}dx^\mu
dx^\nu+dy^2
\nonumber\\
\mbox{4D chosen}\ &:&
H_{\mu\nu\rho\sigma}=\frac{\sqrt{-g}\epsilon_{\mu\nu\rho\sigma}}{
n(y)}\\
&\ &\ H_{5\mu\nu\rho}=0.
\end{eqnarray}
The $H$ field equation is $\partial_M[\sqrt{-g}H^{MNPQ}/H^4]=
\partial_\mu[\sqrt{-g}H^{\mu NPQ}/H^4]=0.$ Thus, we conclude $n$
is a function of $y$ only. For simplicity, we require a $Z_2$
symmetry the symmetry under the reflection: $y\rightarrow -y$. The
bulk solution is\cite{kkl}
\begin{equation}\label{solution}
\beta(|y|)=\left[\frac{(k/a)}{\cosh(4k|y|+c)}\right]^{1/4}
\end{equation}
where $a$ and $c$ are the integration constants, and
\begin{equation}
k=\sqrt{-\frac{\Lambda_b}{6}},\ \ k_1=\frac{\Lambda_1}{6}.
\end{equation}
The boundary condition at $y=0$ is
\begin{equation}
\left.\frac{\beta'}{\beta}\right|_{0^+}=-\frac{\Lambda_1}{6}.
\end{equation}
The bulk solution (\ref{solution}) has the integration constants
$a$ and $c$. $a$ is basically the global charge of the universe
and determines the finite 4D Planck mass. $c$ is the undetermined
integration constant we want and is fixed by the boundary
condition at $y=0$ without fine-tuning, for a finite range of
$|\Lambda_1|<\sqrt{-\Lambda_b/6}$,
\begin{equation}
\tanh c=\frac{k_1}{k}=\frac{\Lambda_1}{\sqrt{-6\Lambda_b}}.
\end{equation}
Since changing $c$ amounts to changing the fields, if the value
$\Lambda_1$ is changed then the fields are expected to adjust so
that the flat space solution results. Note that $\beta(|y|)$ is a
decreasing function of $|y|$ and converges to 0 exponentially as
$|y|\rightarrow\infty$. This property is needed for a self-tuning
solution.\\

The key points of our solution are:\\

\noindent {\bf (1) $\beta$ has no naked singularity--}If we obtain
a 4D flat solution at every bulk point $y$, then the effective 4D
cosmological constant must be zero since there is no singularity.
\\

\noindent {\bf (2) 4D Planck mass is finite--}Even if the extra
dimension is not compact, this theory is permissible since the
gravity is localized near the brane. Integrating with respect to
$y$, we obtain an effective 4D Planck mass
\begin{equation}
M^2_{\rm 4D\ Planck}=\int \sqrt{\beta^4}dy= 2M^3\sqrt{\frac{k}{a}}
\int_0^\infty \frac{1}{\sqrt{\cosh(4ky+c)}}dy
\end{equation}
which is finite and is expected to be of order $M^2$ if the bulk
cosmological constant and the brane tension are of that scale.\\

\noindent {\bf (3) It is a self-tuning solution--}To see that the
4D cosmological constant is zero, we must also consider the
surface term\cite{kkl},
\begin{equation}
S_{\rm surface}=\int d^4xdy\ 4\cdot 4!\left[ \frac{\sqrt{-g}}{H^2}
+A_{NPQ}\partial_M\left(\frac{\sqrt{-g}H^{MNPQ}}{2H^4}\right)
\right].
\end{equation}
Thus, the total action is
\begin{equation}\label{effcc}
S=\int d^4xdy\ \sqrt{-\eta}\beta^4\left[ \frac{R_4}{2\beta^2}
-4\frac{\beta^{\prime\prime}}{\beta}-6
\left(\frac{\beta^\prime}{\beta}\right)^2-\Lambda_b +\frac{2\cdot
4!}{H^2}-\Lambda_1\delta(y) \right]+S_{\rm surface}.
\end{equation}
Then, $-\Lambda_{\rm eff}$ is the $y$ integral of
Eq.~(\ref{effcc}) except the 4D Einstein-Hilbert term $R_4$, to
give vanishing cosmological constant\cite{kkl}. It is consistent
with the original flat 4D ansatz.\\

Therefore, we obtained a flat space solution with $1/H^2$ term.
Certainly, the theory we consider does not belong to the class
studied for the no go theorem, without the Gauss-Bonnet
term\cite{csaki} and with the Gauss-Bonnet term\cite{zee}. [Note
added: After this talk, more self-tuning solutions were found with
a general function of $H^2$ \cite{lee} and with the Gauss-Bonnet
term\cite{binetruy}. With the Gauss-Bonnet term, there is a new
class of solutions which does not include the RS point\cite{GB}.
The solution obtained by Binetruy {\it et al.} belongs to the
class not containing the RS model, or more accurately to the class
containing the Kachru {\it et al.} point\cite{kachru}.]

Note that our solution needs a bulk massless scalar. Namely, only
derivative terms are needed in the bulk, which reminds us of a
Goldstone boson. Another point is that the kinetic energy term is
not a standard one, $1/H^2$. In fact, there are more solutions
with non-standard kinetic energy terms\cite{kkl,lee}, but the
solution we presented is expressed in a closed form and hence is
transparent for discussing the resulting physics.\vskip 0.5cm

\noindent{\bf Dual description}

In the dual picture, we have a 5D scalar $\sigma$. The dual
relation is\cite{ckl}
\begin{equation}
H_{MNPQ}=\sqrt{-g}\epsilon_{MNPQR}\frac{\partial^R\sigma}{
(4!)^{1/3}[(\partial\sigma)^2]^{2/3}}.
\end{equation}
In this dual picture, we obtain the same warp factor with the
self-tuning property. The kinetic energy term of $\sigma$ is
\begin{equation}
-[(\partial\sigma)^2]^{2/3}
\end{equation}
which is not standard. The equation of motion of $H$ becomes the
Bianchi identity in the dual picture, and the Bianchi identity
plays the role of equation of motion in the dual
picture\cite{ckl}. In the dual picture, we can allow the brane
coupling,
\begin{equation}
-\Lambda_1f(\sigma)\delta(y)
\end{equation}
which may be useful for further study of self-tuning solutions.
\vskip 0.5cm

\noindent{\bf De Sitter and anti de Sitter space solutions}
\\

So far we have considered the time-independent solutions. The next
simplest solutions are de Sitter and anti de Sitter space
solutions which are time dependent. For the curvature
$\lambda$($+$ sign for dS and $-$ sign for AdS), the metric is
taken as
\begin{equation}
ds^2=\beta^2(y){\bf g}_{\mu\nu}dx^\mu dx^\nu+dy^2
\end{equation}
where
\begin{eqnarray}
{\bf g}_{\mu\nu}=\left\{\matrix{{\rm
diag.}(-1,e^{2\sqrt{\lambda}t}, e^{2\sqrt{\lambda}t},
e^{2\sqrt{\lambda} t}),\ \ \mbox{for dS}_4 \cr {\rm
diag.}(-e^{2\sqrt{-\lambda}x_3}, e^{2\sqrt{-\lambda}x_3},
e^{2\sqrt{-\lambda} x_3}, 1),\ \ \mbox{for AdS}_4 \cr} \right.
\end{eqnarray}
The 4D Riemann tensor is $R_{\mu\nu}=3\lambda {\bf g}_{\mu\nu}$.
Thus, the (55) and (00) component equations are
\begin{eqnarray}
&6(\beta'/\beta)^2-6\lambda\beta^{-2}=-\Lambda_b-3(\beta^8/A) \\
&3(\beta'/\beta)^2+3(\beta"/\beta)-3\lambda\beta^{-2}=-\Lambda_b
-\Lambda_1\delta(y)-3(\beta^8/A)
\end{eqnarray}
The 4D cosmological constant obtained from the above ansatz should
be $\lambda$. But we cannot show this explicitly since we do not
have an exact solution. Nevertheless, this kind of situation can
be checked from the Karch-Randall(KR) solution with the
fine-tuning ansatz\cite{karch}. For the dS solution,
\begin{equation}
\beta(y)=(\sqrt{\lambda}/k)\sinh[k(y_m-y)],\ \ y_m=(1/k)\coth^{-1}
(k_1/k),
\end{equation}
the effective Planck mass and the effective cosmological constant
are obtained by integrating in $[-y_m,y_m],$
\begin{eqnarray}
&M^2_{P,{\rm eff}}= (\lambda/k^3)(-ky_m+\frac12 \sinh( 2ky_m))>0 \\
&\Lambda_{\rm eff}= (\lambda^2/k^3)(-ky_m+\frac12 \sinh(2ky_m))
=\lambda M^2_{P,{\rm eff}}.
\end{eqnarray}
Therefore, a 4D observer would realize the curvature as $\lambda$.
Similarly, we expect that for our dS and AdS solutions, the 4D
curvature will turn out to be $\lambda$. Our solutions are
depicted in Figs. 5, 6, and 7.

\begin{figure}[htb]
    \centering
    \includegraphics[height=1.5in]{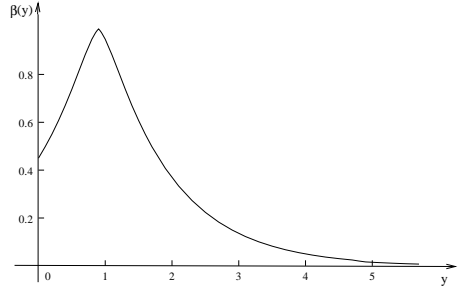}
\caption{The flat space solution.} \label{fig:flat}
\end{figure}
\begin{figure}[htb]
    \centering
    \includegraphics[height=1.5in]{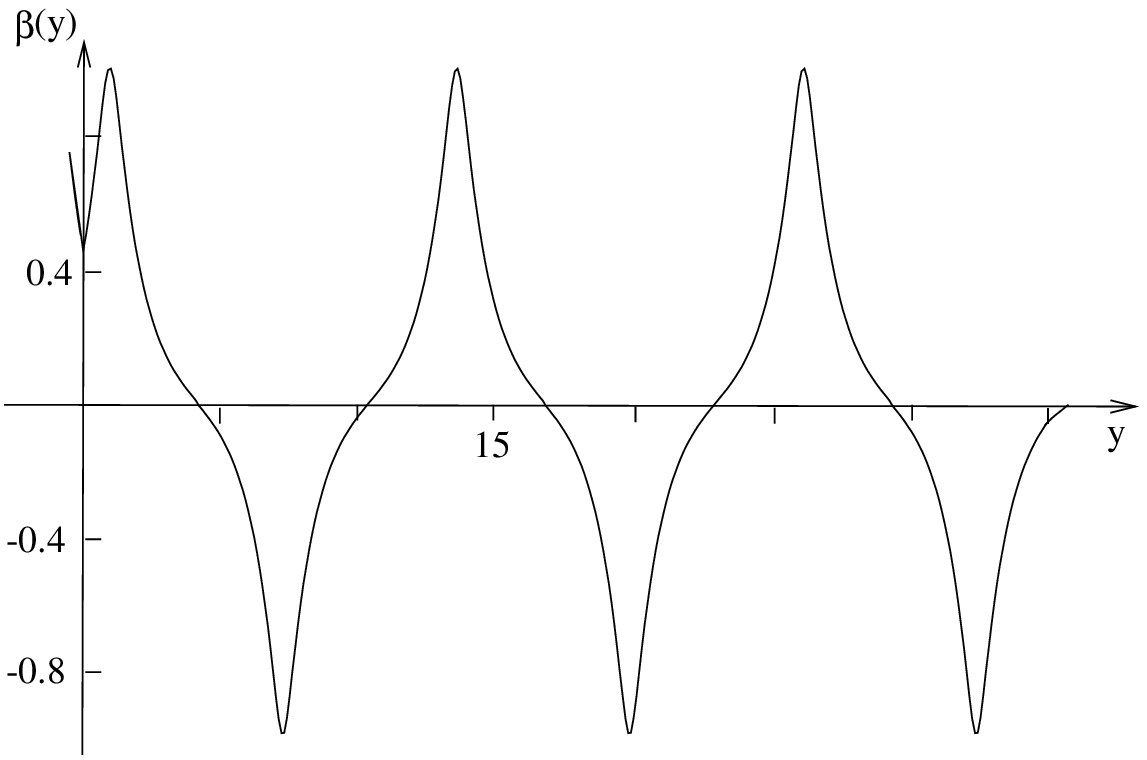}
\caption{The de Sitter space solution.} \label{fig:dS}
\end{figure}
\begin{figure}[htb]
\centering
\includegraphics[height=1.5in]{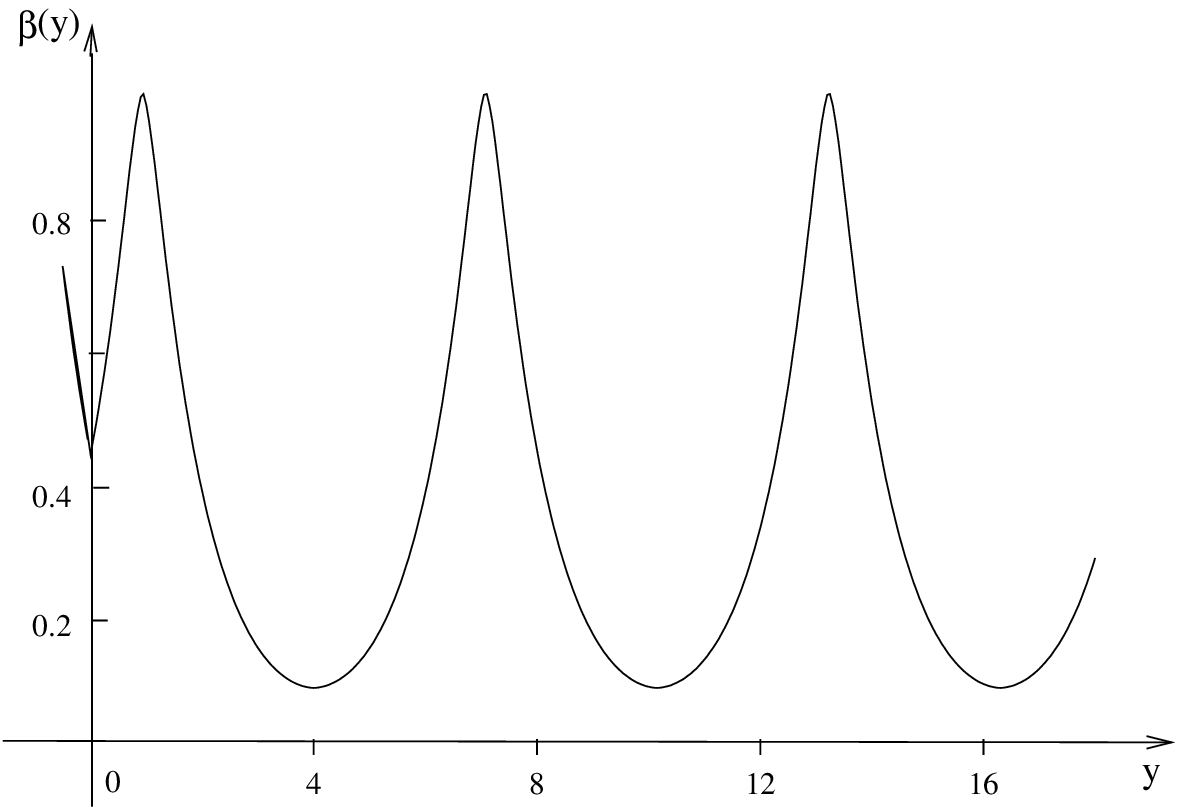}
\caption{The anti de Sitter space solution.} \label{AdS}
\end{figure}

It is easy to observe that these solutions exist. The $y$
derivative of the metric for the dS case is
\begin{equation}
\beta'=\sqrt{k_\lambda^2+k^2\beta^2-a^2\beta^{10}},\ \ {\rm with}\
k_\lambda=\sqrt{\lambda/6}.
\end{equation}
Therefore, for a positive $\lambda$, $\beta'(y_h)$ needs not be
zero at the point where $\beta(y_h)=0$. Thus, there exist a point
$y_h$ where $\beta'$ is finite and $\beta=0$ which is the de
Sitter horizon. It takes an infinite amount of time to reach at
$y_h$. For a negative $\lambda$, $\beta$ can be nonzero where
$\beta'$ is zero. It is an AdS solution, which does not give a
localized gravity. \vskip 0.5cm

\noindent{\bf Probabilistic interpretation}
\\

We note that in 4D with $\Lambda\ne 0$, a flat solution is not
possible. In our case, we obtained flat, dS$_4$, and AdS$_4$
solutions for a finite range of parameters. It is a progress,
since one can choose the flat one if there exists a principle to
do so\cite{hawking,witten}. In particular, let us briefly review
Hawking's idea in 4D. In the Euclidian quantum gravity, the
Euclidian action is
\begin{equation}
-S_E= -\frac12\int d^4x\sqrt{+g}[(R+2\Lambda)+\frac{1}{4!}
H_{\mu\nu\rho\sigma}H^{\mu\nu\rho\sigma}],
\end{equation}
where we used the gravity unit $M^2=1$. The Einstein and $H$
equations are
\begin{eqnarray}
& R_{\mu\nu} -\frac12 g_{\mu\nu}R=\Lambda g_{\mu\nu}-T_{\mu\nu}\\
& \partial_\mu [\sqrt{g}H_{\mu\nu\rho\sigma}]=0
\end{eqnarray}
where
\begin{eqnarray}
H_{\mu\nu\rho\sigma}=\frac{1}{\sqrt{g}}\epsilon_{\mu\nu\rho\sigma}c,\
\ H^2=4!c^2,\ \ T_{\mu\nu}=\frac16 H_{\mu abc} H_{\nu
abc}-\frac{1}{48}g_{\mu\nu}H^2
\end{eqnarray}

\begin{figure}[b]
\begin{center}
\includegraphics[width=.95\textwidth]{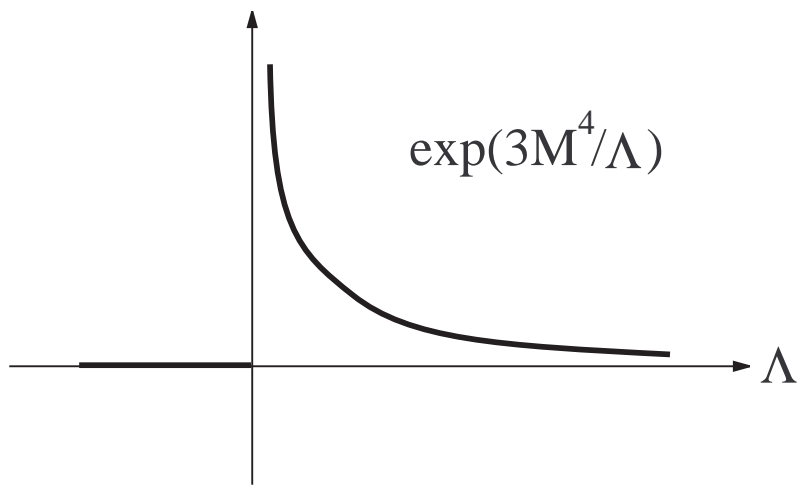}
\end{center}
\caption[]{Hawking's probability function.} \label{eps8}
\end{figure}

Thus, we obtain $T_{\mu\nu}=\frac12 c^2g_{\mu\nu}=-\Lambda_H
g_{\mu\nu}$ where $\Lambda_H=-\frac12 c^2$. The Ricci scalar is
\begin{equation}\label{hawking}
R=-g_{\mu\nu}(R_{\mu\nu}-\frac12 g_{\mu\nu}R)=-4\Lambda_{\rm eff}
\end{equation}
where
\begin{equation}
\Lambda_{eff}=\Lambda+\Lambda_H.
\end{equation}
We obtained (\ref{hawking}) from the equation of motion, but not
from the action itself. If we used the action, then we had to
include the surface term. Therefore, consider for $\Lambda_{\rm
eff}>0$,
\begin{eqnarray}
&-S_E=-\frac12 \int d^4x \sqrt{g}(R+2\Lambda_{\rm eff})=\int d^4x
\sqrt{g}\Lambda_{\rm eff}\nonumber\\
&=\Lambda_{\rm eff}\cdot\mbox{(4D Euclidian
volume)}=\frac{3M^4}{\Lambda_{\rm eff}}
\end{eqnarray}
which is maximum at $\Lambda_{\rm eff}=0^+$, which is shown in
Fig. 8.

Therefore, Hawking interpreted this result as the probability for
the $\Lambda_{\rm eff}=0^+$ universe is maximum in the multi
universe scenario. As explained in the introduction, this 4D
example is not realistic since the universe is originating from
Big Bang where the beginning of time chooses the universe. This
Hawking's argument applies at the beginning. But at later epoches,
the vacuum energies are expected to be added and hence the 4D
example is not complete.

With our self-tuning solution, we can anticipate that one can
choose the flat one in the beginning after inflation and at a
later epoch the flat solution adjusts itself if vacuum energy is
added at the brane.
\\

In conclusion, we presented the interesting self-tuning solutions
of the cosmological constant problem, which has attracted a great
interest recently. This direction seems to be promising in RS-II
type models. We showed an existence proof of such a solution with
the four-form field strength $H$ in 5D. To choose the flat one out
of numerous possibilities, we had to resort to Hawking's
probabilistic interpretation. But there may be some theories where
these strong self-tuning solutions are possible. To find these new
type of solutions, it is worthwhile to observe the key points of
our solutions and presumably one may rely on a symmetry in
theories with a massless scalar coupled to the brane.

\vskip 0.5cm\noindent {\it Acknowledgments}: This work is
supported in part by the BK21 program of Ministry of Education and
by the KOSEF Sundo grant.


\begin{thebibliography}{8.}
\addcontentsline{toc}{section}{References}

\bibitem{gauge} E. Gildener and S. Weinberg, Phys. Rev. {\bf D13},
3333 (1976).

\bibitem{ccp} M. Veltman, Phys. Rev. Lett. {\bf 34}, 777 (1975).
For earlier discussions, refer to Ya. B. Zel'dovich and I. D.
Novikov, {\it Relativistic Astrophysics}, Vol. I (1967 Russian
ed., English Translation in 1971, Univ. of Chicago Press), p.72;
Vol. II (Univ. of Chicago Press, 1983), Chapter 4. For a recent
summary, see, U. Ellwanger, hep-ph/0203252.

\bibitem{strongcp} R. D. Peccei and H. R. Quinn, Phys. Rev. Lett.
{\bf 38}, 1440 (1977).

\bibitem{mu} J. E. Kim and H. P. Nilles, Phys. Lett. {\bf B138},
150 (1984).

\bibitem{weinberg} S. Weinberg, Rev. Mod. Phys. {\bf 61}, 1 (1989).

\bibitem{hawking} S. Hawking, Phys. Lett. {\bf B134}, 403 (1984);
E. Baum, Phys. Lett. {\bf B138}, 185 (1983).

\bibitem{witten} E. Witten, {\it Proc. Shelter Island II
Conference}(Shelter Island, 1983), ed. R. Jackiw, H. Khuri, S.
Weinberg, and E. Witten (MIT Press, 1985), p.369. $H_{MNPQ}$ was
used also by A. Aurila, H. Nicolai, and P. K. Townsend, Nucl.
Phys. {\bf B176}, 509 (1980).

\bibitem{weinberg1} S. Weinberg, Phys. Rev. Lett. {\bf 59}, 2607
(1987).

\bibitem{coleman} S. Coleman, Nucl. Phys. {\bf B310}, 643 (1988).

\bibitem{zel2}
Ya. B. Zel'dovich and I. D. Novikov, {\it Relativistic
Astrophysics}, Vol. I (1967 Russian ed., English Translation in
1971, Univ. of Chicago Press), p.72; Vol. II (Univ. of Chicago
Press, 1983), Chapter 4.

\bibitem{type1a} S. Perlmutter {\it et al.}, The Supernova
Cosmology Project, Bull. Am. Astron. Soc. {\bf 5}, 1351 (1998).

\bibitem{quintessence} For an extremely light light Goldstone
boson as quintessence, see, J. E. Kim, JHEP {\bf 9905}, 022
(1999); JHEP {\bf 0006}, 016 (2000); K. Choi, Phys. Rev. {\bf
D62}, 043509 (2000). For other types, see, P. Binetruy, Phys. Rev.
{\bf D60}, 063502 (1999); C. Kolda and D. H. Lyth, Phys. Lett.
{\bf B458}, 197 (1999); T. Chiba, Phys. Rev. {\bf D60}, 083508
(1999); P. Brax and J. Martin, Phys. Lett. {\bf B468}, 40 (1999);
A. Masiero, M. Pietroni, and F. Rosati, Phys. Rev. {\bf D61},
023504 (2000); M. C. Bento and O. Bertolami, Gen. Rel. Grav. {\bf
31}, 1461 (1999); F. Perrotta, C. Baccigalupi, and M. Matarrese,
Phys. Rev. {\bf D61}, 023507 (2000).

\bibitem{rs1} L. Randall and R. Sundrum, Phys. Rev. Lett. {\bf
83}, 3370 (1999).

\bibitem{rs2} L. Randall and R. Sundrum, Phys. Rev. Lett. {\bf
83}, 4690 (1999).

\bibitem{kachru} N. Arkani-Hamed, S. Dimopoulos, N. Kaloper, and
R. Sundrum, Phys. Lett. {\bf B480}, 193 (2000); S. Kachru, M.
Schulz, and E. Silverstein, Phys. Rev. {\bf D62}, 045021 (2000).

\bibitem{GB} J. E. Kim, H. M. Lee, and B. Kyae, Phys. Rev. {\bf D62},
045013 (2000); Nucl. Phys. {\bf B582}, 296 (2000).

\bibitem{nilles} S. F$\ddot{\rm o}$rste, Z. Lalak, S. Lavignac,
and H. P. Nilles, JHEP {\bf 0009}, 034 (2000).

\bibitem{lee1} J. E. Kim, B. Kyae, and H. M. Lee, hep-th/0110103.

\bibitem{kkl} J. E. Kim, B. Kyae, and H. M. Lee, Phys. Rev. Lett.
{\bf 86}, 4223 (1991); Nucl. Phys. {\bf B613}, 306 (2001).

\bibitem{csaki} C. Csaki, J. Erlich, C. Grojean, and T. Hollowood,
Nucl. Phys. {\bf B584}, 359 (2001).

\bibitem{zee} I. Low and A. Zee, Nucl. Phys. {\bf B585}, 395 (2000).

\bibitem{lee} J. E. Kim and H. M. Lee, hep-th/0207260.

\bibitem{binetruy} P. Binetruy, C. Charmousis, S. C. Davis, and
J.-F. Dufaux, hep-th/0206089.

\bibitem{ckl} K.-S. Choi, J. E. Kim, and H. M. Lee, J. Korean
Phys. Soc., {\bf 40}, 207 (2002) [hep-th/0201055].

\bibitem{karch} A. Karch and L. Randall, JHEP {\bf 0105}, 008 (2001).

\end{thebibliography}
\end{document}